\documentclass[useAMS,usenatbib]{mn2e}
\usepackage[colorlinks]{hyperref}
\usepackage{amssymb}
\usepackage{amsmath}
\usepackage{txfonts}
\usepackage{graphicx}
\usepackage{subcaption} 

\usepackage{epsfig}
\usepackage{xspace}
\usepackage{url}
\usepackage[colorinlistoftodos]{todonotes}

\usepackage{natbib}
\def\swift{\textit{Swift}\xspace}

\def\cha{\textit{Chandra}\xspace}

\def\xrt{\textit{Swift}/XRT\xspace}

\def\flat{\textit{Fermi/LAT}\xspace}

\def\psrb{PSR~B1259-63\xspace}

\def\psrbls{PSR~B1259-63/LS~2883\xspace}

\title[Origin of the GeV flare in  PSR B1259-63]{New insight into the origin of the GeV flare in  the binary system PSR B1259-63 / LS 2883 from the 2017 periastron passage.}
\begin{document}
\author[M. Chernyakova et al.]{M. Chernyakova,$^{1,2}$\thanks{E-mail: masha.chernyakova@dcu.ie} 
          D. Malyshev,$^{3}$ 
          S.  Mc Keague,$^{1}$ 
          B. van Soelen,$^{4}$  
          J.P. Marais,$^{4}$ 
          \newauthor
          A. Martin-Carrillo,$^{5}$ 
          and D. Murphy$^5$
          \\
$^{1}$ School of Physical Sciences and CfAR, Dublin City University, Dublin 9, Ireland\\
$^{2}$ Dublin Institute for Advanced Studies, 31 Fitzwilliam Place, Dublin 2, Ireland\\
$^{3}$ Institut f{\"u}r Astronomie und Astrophysik T{\"u}bingen, Universit{\"a}t T{\"u}bingen, Sand 1, D-72076 T{\"u}bingen, Germany \\
$^{4}$ University of the Free State Department of Physics PO Box 339 9300 Bloemfontein South Africa\\
$^{5}$ Space Science Group, School of Physics, University College Dublin, Belfield, Dublin 4, Ireland
}
\date{Received $<$date$>$  ; in original form  $<$date$>$ }

\label{firstpage}
\pagerange{\pageref{firstpage}--\pageref{lastpage}} 
\pubyear{2020}

\maketitle

\begin{abstract}
\psrb is a gamma-ray binary system hosting a  radio pulsar orbiting around a O9.5Ve  star, LS 2883,  with a period of $\sim$ 3.4 years. The interaction of the pulsar wind with the LS 2883 outflow leads to unpulsed broad band emission in the radio, X-rays, GeV and  TeV domains. While the radio, X-ray and TeV light curves show rather similar behaviour, the GeV  light curve appears very different with a huge outburst about a month after a periastron. The energy release during this outburst seems to significantly exceed the spin down luminosity of the pulsar and both the GeV light curve and the energy release varies from one orbit to the next. 

In this paper, we present for the first time the results of optical observations of the system in 2017, and also reanalyze the available X-ray and GeV data. We present a new model in which the GeV data are explained as a combination of the bremsstrahlung and inverse Compton emission from the unshocked and weakly shocked electrons of the pulsar wind. The X-ray and TeV emission is produced by synchrotron and inverse Compton emission of energetic electrons accelerated on a strong shock arising due to stellar/pulsar winds collision. The brightness of the GeV flare is explained in our model as a beaming effect of the energy released in a cone oriented, during the time of the flare, in the direction of the observer.

\end{abstract}

\section{Introduction}
Gamma-ray binary systems are composed of a compact object, either a black hole or a neutron star, orbiting a massive O or B type star. They are distinguished from X-ray binaries of a similar nature by non-thermal emission that peaks at energies 
$\gtrsim 1$~MeV \citep{2017A&A...608A..59D}.
Of all $\gamma$-ray binaries the only systems where the nature of the compact object is known are \psrb and PSR J2032+4127, both of which are radio pulsars. In \psrb the pulsar has a spin period of 47.76~ms and is orbiting a O9.5Ve  star (LS~2883)  with a period of $\sim 1236.7$~days  in a highly eccentric orbit ($e\sim 0.87$) \citep{1992ApJ...387L..37J,Negueruela-2001,shannon14}.  Based on the parallax data in the Gaia DR2 Archive~\citep{Gaia2018} the distance to the system is $2.39 \pm 0.19$~kpc, which is consistent with the value of $2.6^{+0.4}_{-0.3}$ kpc is reported by \cite{PSRB1259-2018_distance}.

The optical spectrum of the companion
shows evidence of an equatorial disc, which is thought to be inclined with
respect to the orbital plane  by $\sim 10-40^\circ$ \citep{Melatos95}, which causes the pulsar to cross the disc twice during the periastron passage. 
The interaction of the pulsar wind  with the companion's outflow leads to the generation of the  unpulsed non-thermal emission in radio, X-ray, GeV and TeV energies. X-ray emission is observed throughout the orbit but the unpulsed radio, GeV and TeV radiation occurs only within a few months before and after the periastron \citep[e.g.][]{1999MNRAS.302..277J, chernyakova15, 2018ApJ...863...27J,hess_psrb2020}.

The unpulsed radio and X-ray emission exhibits a similar two peak light curve with the peaks occurring during the time when the pulsar crosses the disc of the companion. Current H.E.S.S. observations indicate that TeV emission can have a similar behaviour \citep{hess_psrb2020}, but more sensitive observations are needed to confirm this. Hopefully CTA will address this issue in the very near future \citep{CTArev2019}. The GeV emission, however, shows a very different behaviour and is characterised by a strong flare, which started about $\sim 30$~days after the 2010 and 2014 periastron passages \citep{2011ApJ...736L..11A,caliandro15} and has no obvious flaring counterparts at other wavelength. 

The only visible effect coinciding in time with a GeV flare is a rapid decrease of the H\,$\alpha$ equivalent width \citep{chernyakova15}, usually interpreted as a measure of the companion's star disc. The destruction of the disc is also evident in Chandra observations of the source far away from the periastron \citep{Pavlov2011,Pavlov2015,Pavlov2019,Hare2019}. These data demonstrates the presence of X-ray emitting clumps moving away from the binary with speeds of about 0.1 of the speed of light. The clumps are being ejected at least once per binary period, 3.4 years, presumably around binary periastra and are probably associated with the destruction of the companion's disc.

The most recent periastron of \psrb  (September 22, 2017; $t_p = \mbox{ MJD~} 58018.1$)  presents another opportunity to examine the nature and mechanics of the GeV flare using the available broadband observations.  The GeV behaviour of \psrb turned out to be very different from previous periastra. The flare only started 40 days after the periastron and reveals variability on hour timescales \citep[e.g.][]{2018ApJ...863...27J,Tam2018,Ghang2018}. 

In this paper we present for the first time the results of the optical observations of the system in 2017, discuss the available multiwavelength data (optical, X-ray and GeV) and propose a new model to explain the origin of the GeV flare and how its observed luminosity apparently exceeds the spin-down luminosity. In section~\ref{sec:data_analysis} of this paper we describe the details of the data analysis, in section~\ref{sec:discussion} we present our model, and give our conclusions in section~\ref{sec:conclusions}.

\section{Data analysis}
\label{sec:data_analysis}
\subsection{Optical observations and analysis }

Optical observations during the 2017 periastron passage were limited because of the position of the source relative to the Sun, and \psrbls was only visible for a short period just after sunset, allowing limited observations before periastron. 
The system was observed with the SAAO 1.9-m telescope between 2017 August 24 and  2017 September 04  ($\tau \sim -28$\,d to $\tau\sim -17$\,d; where $\tau$ is the time from periastron) using the SpUpNIC grating spectrograph \citep{2016SPIE.9908E..27C}.  

The spectroscopic observations with SpUpNIC were performed using a 1200 lines\,mm$^{-1}$ grating, with a spectral resolution of  $\sim 1$\,\AA{}, covering a wavelength range of $\sim 6150 - 7150$\,\AA{}.  Each night, multiple exposures of the target were taken with a typical exposure time of 60 seconds, while the source was high enough to be observed. Arc observations of a CuNe arc lamp were taken before and after every science exposure.  Dome flats where taken each day using the same configuration and a spectroscopic standard (CD-32 9927) was observed at the beginning of each night.  The data reduction and wavelength calibration was perform following the standard {\sc iraf/noao} procedures. Each night's observations were combined and normalized.

As previously reported, the spectrum shows a strong H\,$\alpha$ emission line that remains single peaked through all observations.
The double peaked He\,{\sc i} ($\lambda$6678) line is also present in the observations and the variation of the ratio of the peaks of the violet to red  (V/R) components  of the line was measured.  

Further photometric observations in the H\,$\alpha$ filter were undertaken using the Watcher Robotic Telescope \citep{french04} from 2017 September 04 to 2017 September 21 ($\tau \sim -17$\, to $\sim -0.4$\,d). Multiple 30\,second exposure in the H\,$\alpha$ filter were taken per night.  All images on the same night were combined to increase the signal to noise and differential photometry, using stars on the same field of view, was performed.  

\begin{figure}
\includegraphics[width=\linewidth]{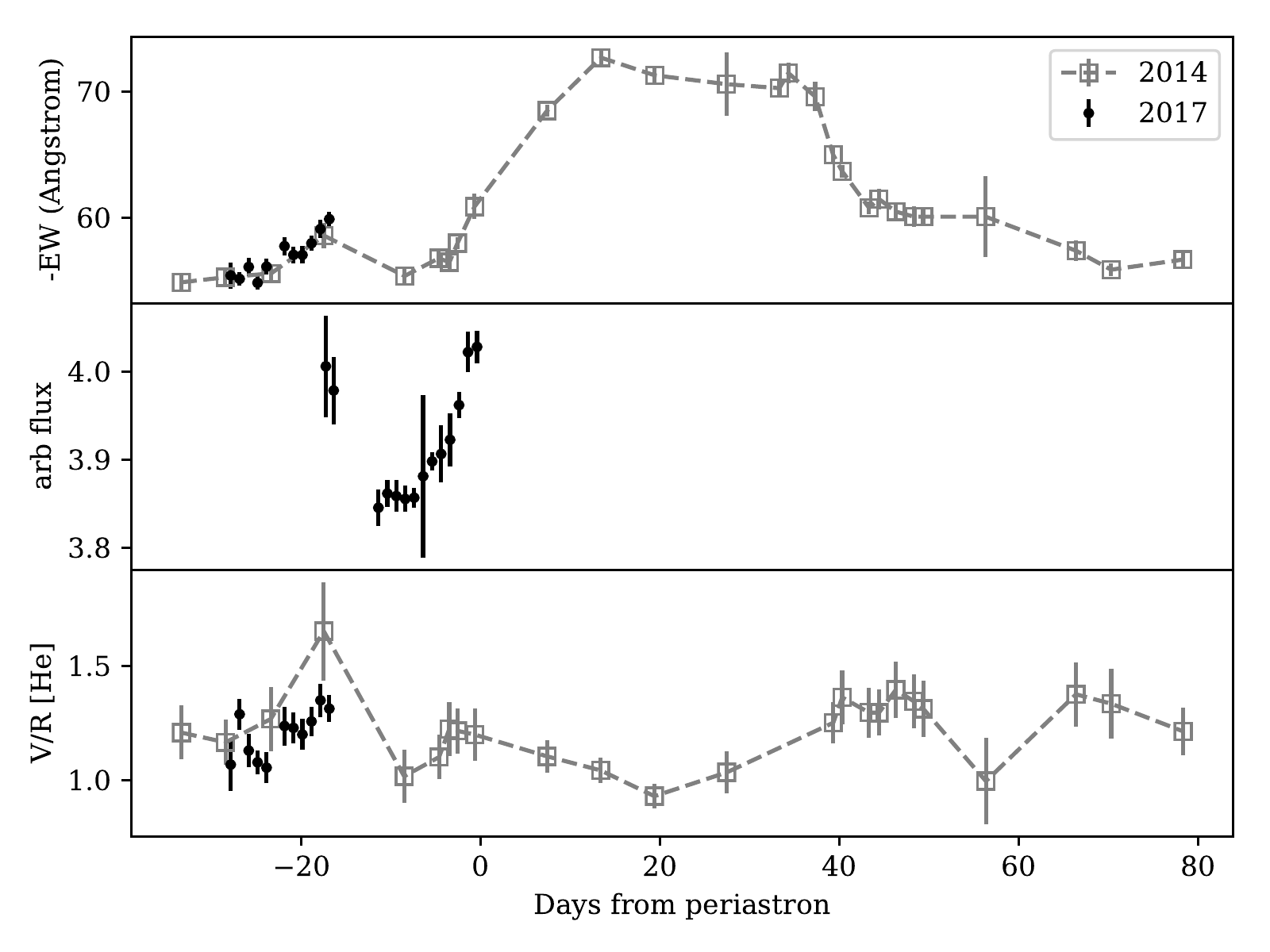}
\caption{Optical observations before the 2017 periastron passage. {\bf Top:} H\,$\alpha$ equivalent width. {\bf Middle:} H\,$\alpha$ photometry (in arbitrary units) {\bf Bottom:} V/R line ratio from the He I profile. The results are compared to the 2014 periastron passage.}
\label{fig:optical}
\end{figure}

The results are shown in Fig.~\ref{fig:optical} (and also bottom panel of Fig.~\ref{fig:varper}) and compared to the observations around the 2014 periastron passage \citep{2014MNRAS.439..432C,2016MNRAS.455.3674V}. The top panel shows that the H\,$\alpha$ equivalent width follows the same trend as the previous periastron passage, with the line strength increasing towards the point of the first disc crossing. While the spectroscopic observations could not 
be taken beyond this, the photometric H\,$\alpha$ observations (middle panel, shown in arbitrary flux units) show that the line continues to follow the same trend; the line strength decreases after the first disc crossing, then continues to grow towards periastron.  This confirms the strong interaction between the pulsar and the circumstellar disc around the periastron. This is also shown by the V/R variation (bottom panel) which follows a similar trend, though at a lower scale, with the V component increasing towards the first disc crossing.  The optical spectroscopic results are given in Table.~\ref{tab:optical}.

\begin{table}
 \caption{The H\,$\alpha$ equivalent width and V/R ratio of the double peaked He\,{\sc i} line measured around the first disc crossing phase before the 2017 periastron passage. }
 \label{tab:optical}
  \begin{tabular}{lcc} \hline
Days from periastron & H\,$\alpha$ equivalent width ($\AA$)   & He\,{\sc i} V/R   \\ \hline
-27.90 & $ -55.4 \pm 1.0 $ & $ 1.07 \pm 0.11 $ \\
-26.90 & $ -55.2 \pm 0.5 $ & $ 1.29 \pm 0.07 $ \\
-25.89 & $ -56.1 \pm 0.7 $ & $ 1.13 \pm 0.07 $ \\
-24.90 & $ -54.9 \pm 0.5 $ & $ 1.08 \pm 0.05 $ \\
-23.90 & $ -56.1 \pm 0.7 $ & $ 1.05 \pm 0.07 $ \\
-21.90 & $ -57.8 \pm 0.7 $ & $ 1.24 \pm 0.08 $ \\
-20.90 & $ -57.1 \pm 0.6 $ & $ 1.23 \pm 0.07 $ \\
-19.90 & $ -57.1 \pm 0.7 $ & $ 1.20 \pm 0.07 $ \\
-18.89 & $ -58.0 \pm 0.6 $ & $ 1.26 \pm 0.06 $ \\
-17.88 & $ -59.1 \pm 0.7 $ & $ 1.35 \pm 0.07 $ \\
-16.90 & $ -59.9 \pm 0.6 $ & $ 1.31 \pm 0.06 $ \\ \hline
\end{tabular}
\end{table}

\subsection{X-ray data}

A full overview of the X-ray flux and spectral slope for the observations around  2004, 2007, 2010, 2014 and 2017 years are presented in the panels (b) and (c) of Fig.~\ref{fig:varper}. Historical data in this figure are taken from 
\citet{chernyakova15}

\subsubsection{\xrt}
The 2017 periastron passage of \psrb was closely monitored by the \swift satellite \citep{2004ApJ...611.1005G}. We have analysed all available data taken from March, 26th, 2017 to January, 6th, 2018. 

The data were reprocessed and analysed as suggested by the \xrt team\footnote{See e.g. the \href{https://swift.gsfc.nasa.gov/analysis/xrt_swguide_v1_2.pdf}{\xrt User's Guide}} with the \texttt{xrtpipeline v.0.13.5} and \texttt{heasoft v.6.27} software package. The spectral analysis of \xrt spectra was performed with \texttt{XSPEC v.12.11.0}. The spectrum was extracted from a circle of radius $36''$ around the position of \psrb and the background estimated from a co-centred annulus with inner/outer radii of $60''/300''$.

During the spectral analysis we noticed that the quality of the \xrt 
data does not allow  the hydrogen column density to be firmly determined in each individual observation and thus we chose to fix it to the mean value. To do this we have fitted all the data with an absorbed power law model with a common value of column density in all the observations. The resulting value of $N_H=0.55\times10^{22}$cm$^{-2}$ is in good agreement with previous observations \citep[see e.g.][and references therein]{chernyakova15}. These data were also presented in the paper of \citet{Tam2018}, but in that work the value of the column density was fixed in a rather model-dependent way.

\subsubsection{\cha}
We accompanied our analysis with the analysis of historic, publicly available \cha data taken during the 2014 and 2017 periastron passages (February to June 2014, ObsIds: 16563, 16583, 16624, 16625 and July 2017, ObsIds: 19281,20116).

We analyzed these data using the most recent \texttt{CIAO v.4.12} software and CALDB 4.9.0. The data were reprocessed with the \texttt{chandra\_repro} utility. The source and background spectra, with corresponding RMFs and ARFs, were extracted with the \texttt{specextract} tool. We note that two \cha observations of \psrb in June 2014 (ObsIds: 16624, 16625) were performed in asic-cc (continuous clocking) mode, in which only 1-dimensional spatial information is available. In this case, to extract the source and background spectra, we used box-shaped regions\footnote{See \href{http://cxc.harvard.edu/ciao/caveats/acis_cc_mode.html}{caveats of asic-cc mode data analysis.}} centred on \psrb and on a nearby source-free region, respectively. For the rest of observations we utilized standard circular regions.

\begin{figure}
\includegraphics[width=\columnwidth]{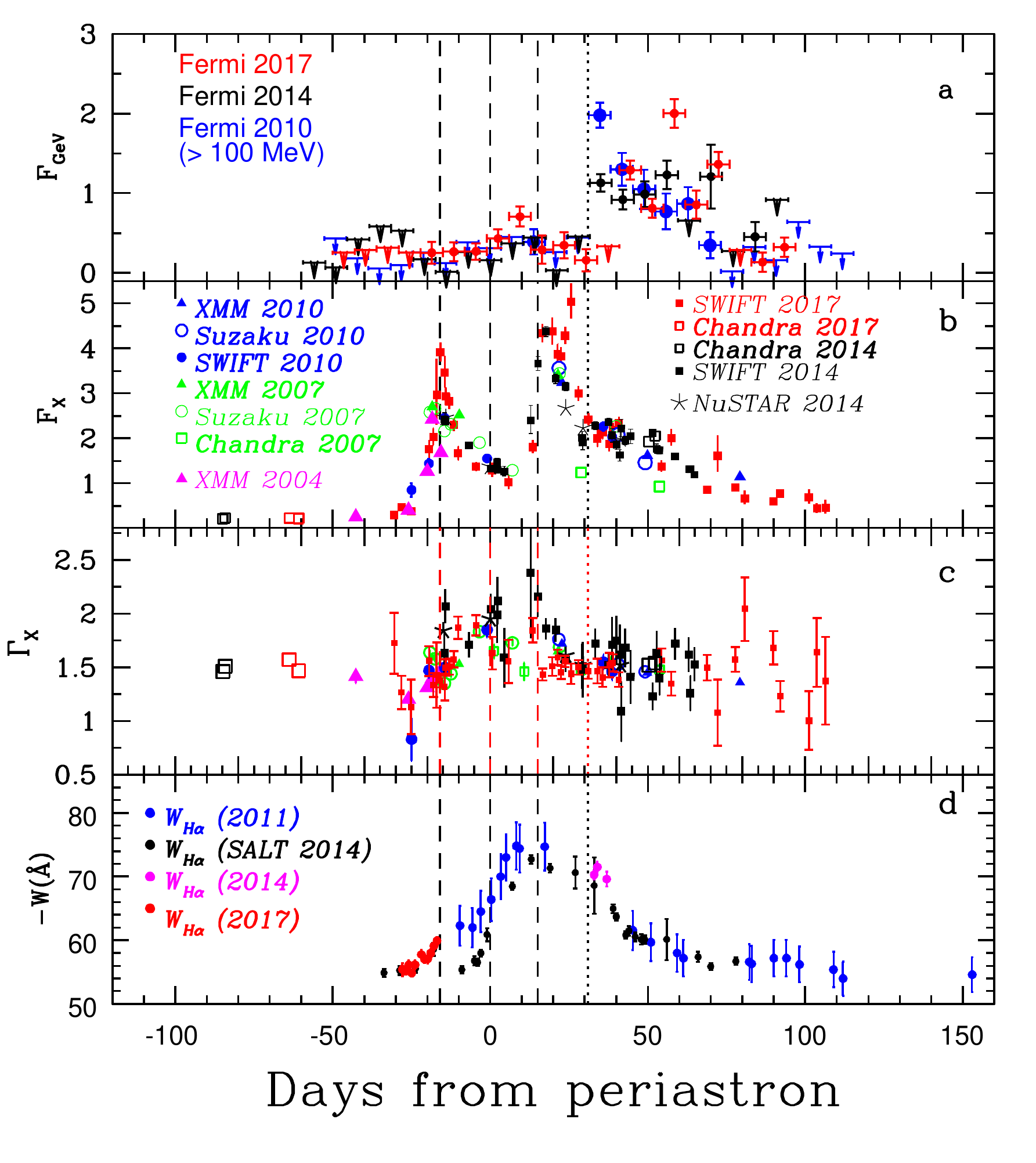}
\vspace{-1cm}
\caption{Evolution of multiwavelength \psrb spectral characteristics over the different periastron passages. \textit{Panel a:} \flat flux measurements in the E $>$ 100 MeV energy range with a weekly bin size. Flux is given in 10$^{-6}$ cm $^{-2}$ s $^{-1}$. \textit{Panel b:} 1-10 keV X-ray flux in units of 10$^{-11}$ erg cm $^{-2}$ s $^{-1}$.  \textit{Panel c:} X-ray slope. \textit{Panel d:} H\,$\alpha$ equivalent width. 
 }
\label{fig:varper}
\end{figure}

\subsection{\flat Observations and Analysis}
The analysis of \flat data was performed using Fermitools version 1.2.23 (released 11th February 2020). For the analysis of the 2017 periastron passage and the combined periastron data the analysis was carried out using the latest Pass 8 reprocessed data (P8R3) from the SOURCE event class. All gamma-ray photons used for this analysis were within the energy range 0.1 -- 100~GeV and within a circular region of $15^\circ$ around the ROI centred on \psrb. The selected maximum zenith angle was $90^\circ$. The spatial-spectral model built in order to perform the likelihood analysis included the Galactic and isotropic diffuse emission components and the known gamma-ray sources within $20^\circ$ of the ROI centre from the 4FGL catalogue \citep{4FGL}. A likelihood analysis was applied to each observation data set twice to achieve the best model for that time span and energy range. The first likelihood run frees the normalization of every source within $15^\circ$ of the ROI centre and the index of \psrb. The second run fixes the normalization of all sources outside $5^\circ$ of \psrb to the value calculated from the first likelihood fit. The output model of 
this second likelihood analysis was used for the lightcurve and spectrum generation.

The gamma-ray flux, light-curves and spectral results for PSR B1259-63 presented here were calculated using a binned likelihood fit using the BinnedAnalysis module from the FermiTools Python packages. 
\psrb was modelled using a single powerlaw with the normalization left free and the index was fixed to the value from the preliminary analysis. 
When generating the light curves and spectra, any free sources (except \psrb) with a TS $<$ 1 was removed from the fitted model. If any bin had a poor detection of PSR B1259-63 (TS $<$ 1 or Flux $<$ Flux Error) the calculated flux was replaced with a 95\% confidence upper limit on the photon flux above 100 MeV using the IntegralUpperLimits functions from gtlike. The spectral model output of the poor detection fit was used to calculate the upper limit values. 

Data covering a time frame from 50 days before the periastron to 100~days after periastron were used to produce lightcurves of the 2017 periastron passage. The results of the lightcurve analysis were used to define the time periods for further spectral analysis. 

\begin{figure}
\includegraphics[width=1.\linewidth]{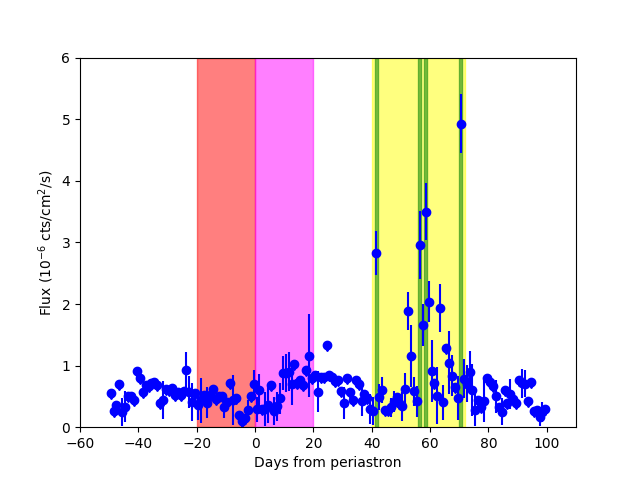}

\caption{Daily-binned light curve of the 2017 periastron passage. The highlights indicate the time periods that were used for spectral analysis and modelling; the details are given in Table \ref{tab_peri}. The different periods are shown as: prfl1 = Red, prfl2 = Magenta, avfl = Yellow, pkfl = Green.}
\label{fig:day_LC}
\end{figure}

As seen in the top panel of Fig. \ref{fig:varper}, the weekly binned GeV light curves of the 2010, 2014 and 2017 periastra show differences in the shape of the post-periastron flaring period. The 2017 GeV light curve, with daily binning, is shown in Fig. \ref{fig:day_LC}. This light curve demonstrates strong day to day flux variability. The different colour highlights on this figure are used to show the data sets used for spectral analysis (see also Table \ref{tab_peri}).
The results of this light curve analysis are in line with previous analyses of the 2017 periastron passage by \citet{Tam2018} and \citet{2018ApJ...863...27J} where discrepancies are likely caused by the use of different catalogues and updated software.

\begin{figure}
\includegraphics[width=\linewidth]{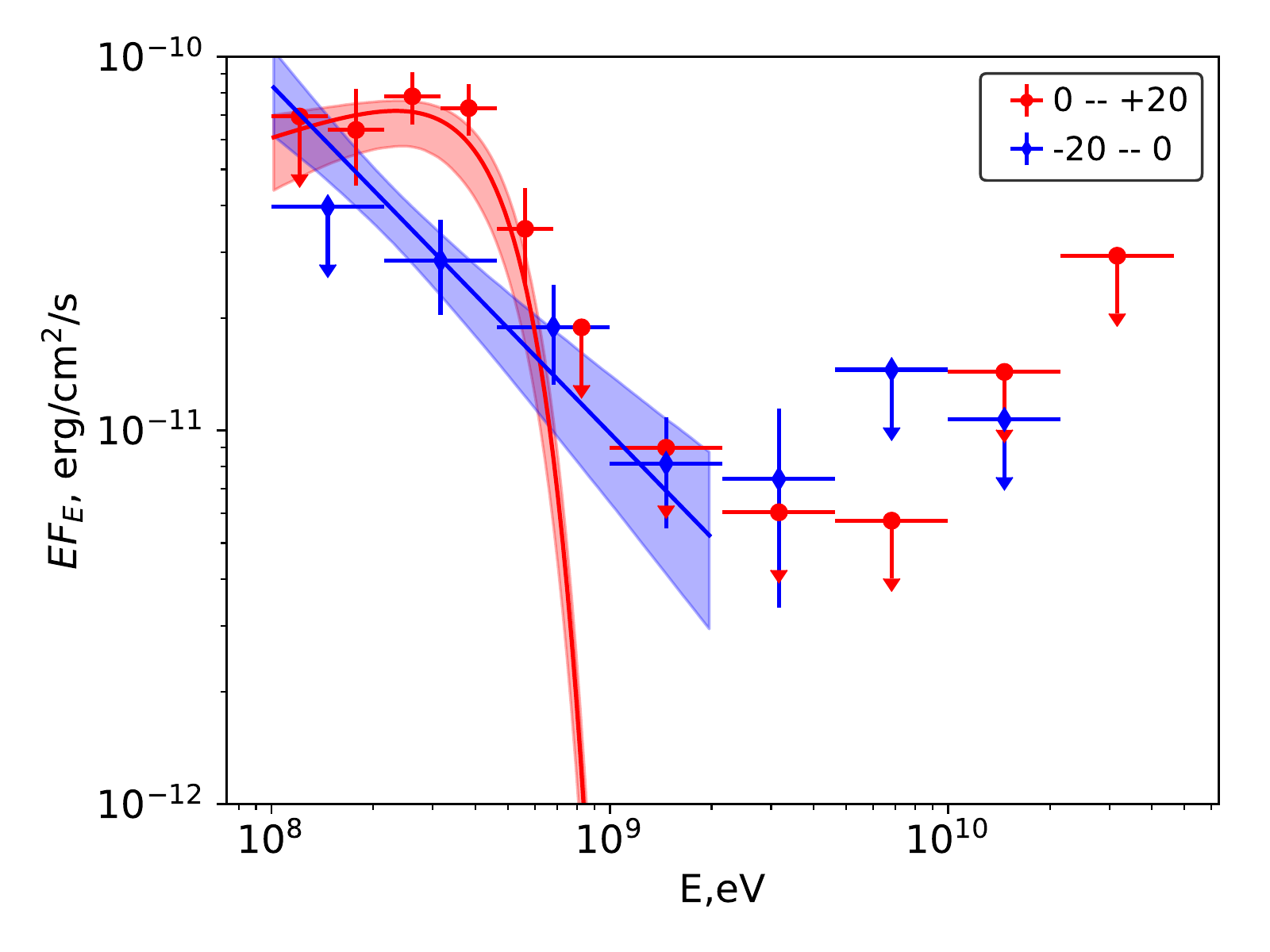}
\caption{GeV emission of \psrb during the periods twenty days before (blue points, prfl1) and after (red points, prfl2) the periastron. Blue and red curves show best fit models in 100 MeV -- 2 GeV energy range, see Table \ref{tab_flare}. Shaded regions show 1 $\sigma$ confidence range for fitted models.}
\label{gev_preflare}
\end{figure}

For the spectral analysis we split the available data into several 
time periods (see Table \ref{tab_peri}). 
The period before the GeV flare (pre-flare), was divided into two: from  20 days before until periastron, and from periastron until  20 days after (prfl1 and prfl2 data sets correspondingly).
For these periods data from the 2010, 2014 and 2017 periastra were included to improve the statistics. 
For the flare analysis we use only data from the 2017 periastron to examine the spectra of the average flare period and the daily short flares that can be seen in the Fig. \ref{fig:day_LC}.
The pre-flare spectra before and after the periastron are compared in Fig. \ref{gev_preflare} and, along with the flare spectra, are also shown in Fig. \ref{fig:avflare}. All time periods defined in Table \ref{tab_peri} were also analysed 
in the $0.1 - 2.0$~GeV energy range. We  used a super exponential  cut-off power law (PLSuperExpCutoff) model to match the shape of the GeV peak spectra;
see Table \ref{tab_flare} for the best-fit parameters. Spectral fitting was done twice, the first time with all the parameters free, and second time with 
$\gamma2$ fixed to the value of the first fit to better constrain the other model parameters.

\begin{table}
\caption{Details of the time periods used for spectral analysis during the 2017 periastron. $t_p=\mbox{MJD } 58018.1$ corresponds to the time of 2017 periastron passage. } 
\label{tab_peri}
\begin{center}
\begin{small}
	\begin{tabular}{|c|c|c|c|}
 \hline
	Data Set  & Time period (MJD)  & t-t$_p$ (days)   \\ \hline
	prfl1   & 55524.7 -- 55544.7,      & -20 -- 0   \\ 
	& 56761.4 -- 56781.4, && \\	
	& 57998.1 -- 58018.1 && \\ \hline
	prfl2   & 55544.7 -- 55564.7,       & 0 -- 20     \\ 
	& 56781.4 -- 56801.4, && \\
	& 58018.1 -- 58038.1 && \\ \hline
    avfl	& 58058.1 -- 58090.1     & 40 -- 72 	\\ \hline
    pkfl	& 58059.1, 58074.1, 58076.1, 58088.1     & 41, 56, 58, 70 	\\ \hline
    \end{tabular}
\end{small}
\end{center}
\end{table}

\begin{table}
\caption{Spectral model parameters from the binned likelihood analysis of different phases of the 2017 periastron from 0.1 - 2.0 GeV. The model used is the PLSuperExpCutoff where $dN/dE = N_0 (E/E_c)^{\gamma1} exp(-(E/E_{0})^{\gamma2})$.}
\label{tab_flare}
\begin{center}
\begin{small}
	\begin{tabular}{|c||c|c|c|c|}
 \hline
	Data Set 	&  $\gamma1$	& $E_c$   &  $\gamma2$   & Photon Flux  \\     
	&& (GeV) &&(10$^{-6}$ cts/cm$^{2}$/s)\\ \hline
	prfl1   & $-2.9 \pm 0.3$    & $-$     & $-$     & $0.26 \pm 0.05$  \\ \hline
	prfl2   & $-1.5 \pm 0.4$    & $0.50 \pm 0.08$   & $3.0$     & $0.28 \pm 0.04$  \\ \hline
	avprfl 	& $-2.4 \pm 0.3$ 	& $0.65 \pm 0.10$ 	& $5.0$     & $0.34 \pm 0.05$ 	\\ \hline
    avfl 	& $-2.4 \pm 0.1$ 	& $1.53 \pm 0.23$	& $3.0$     & $1.11 \pm 0.09$	\\ \hline
    pkfl 	& $-2.3 \pm 0.1$ 	& $1.86 \pm 0.43$	& $3.0$     & $3.39 \pm 0.38$	\\ \hline
    \end{tabular}
\end{small}
\end{center}
\end{table}

The spectral shape of the GeV emission turned out to be very different before and after the periastron, see Figure \ref{gev_preflare}. Blue and red points correspond to the spectrum of the source averaged over 20 days before and after the periastron correspondingly. The spectrum before periastron (prfl1) is well described with a simple power law. The GeV emission after the periastron (prfl2) has a higher flux in the 0.1 -- 2~GeV energy range and is characterised by a strong cut-off, see Table \ref{tab_flare}. 
This noticeable difference in spectral characteristics between different time periods of the preflare implies that around periastron there is a change of the spectrum of electron population responsible for the GeV production.

While there is a notable distinction in the characteristics between the time periods of the preflare, such a difference was not observed in the analysis of the flare periods. The spectra of both the average flare period (avfl) and the peaks of the flare (pkfl) are quite similar (see Figure \ref{fig:avflare} and Table \ref{tab_flare}), and differ mainly by the normalisation.

\section{Discussion}
\label{sec:discussion}
The energy release in \psrb is known to be extremely efficient, for example around periastron more than  10\% of the spin-down luminosity is released in the $1 - 10$~keV energy band alone \citep[e.g.][]{Chernyakova2017}. The energy release during the GeV flare is even larger, reaching values higher than 50\% of the spin-down luminosity in 2010 and 2014 years, without considering possible beaming effects \citep{caliandro15}. The GeV flare in 2017 differs a lot from the previous ones, as it starts about 10 days later and consists of a number of short individual flares, more intensive than previously observed \citep{Tam2018,2018ApJ...863...27J}. The brightness of these flares allows their structure to be reconstructed on timescales shorter than 1 day, and even to find 15 minute long sub-flares. The energy release during these sub-flares reached a value exceeding the spin-down luminosity by a factor of 30, with no clear counterparts at other wavelength~\citep{2018ApJ...863...27J}. This makes it clear that the GeV flare is produced as a separate and highly anisotropic component.

 In what follows we propose a new model, which suggests the presence of two populations of relativistic electrons: \textit{(i):} electrons of the unshocked and weakly shocked pulsar wind and \textit{(ii):} strongly shocked electrons. 
 
 The spectrum of unshocked electrons was selected to be a power law with the slope $-2$ in energy range $E_e=0.6-1$~GeV.  A small fraction of electrons are additionally accelerated at the strong shock near the apex  to $E_e\sim 500$~TeV energies with the similar slope $\Gamma_e=-2$ on a characteristic timescale (see Fig.~\ref{fig:losses}) 
\begin{equation}
t_{\rm acc} \approx 0.1\left(E_e/1\,\mbox{TeV}\right)\eta (B_0/1\,\mbox{G})^{-1}\quad\mbox{s}
\label{eq:t_acc}    
\end{equation}
where $B_0$ is a magnetic field in the region and $\eta\geq 1$ is the acceleration efficiency~\citep[see e.g.][for the details]{LS5039_2008_Khan}.
 
\begin{figure}
\includegraphics[width=\linewidth]{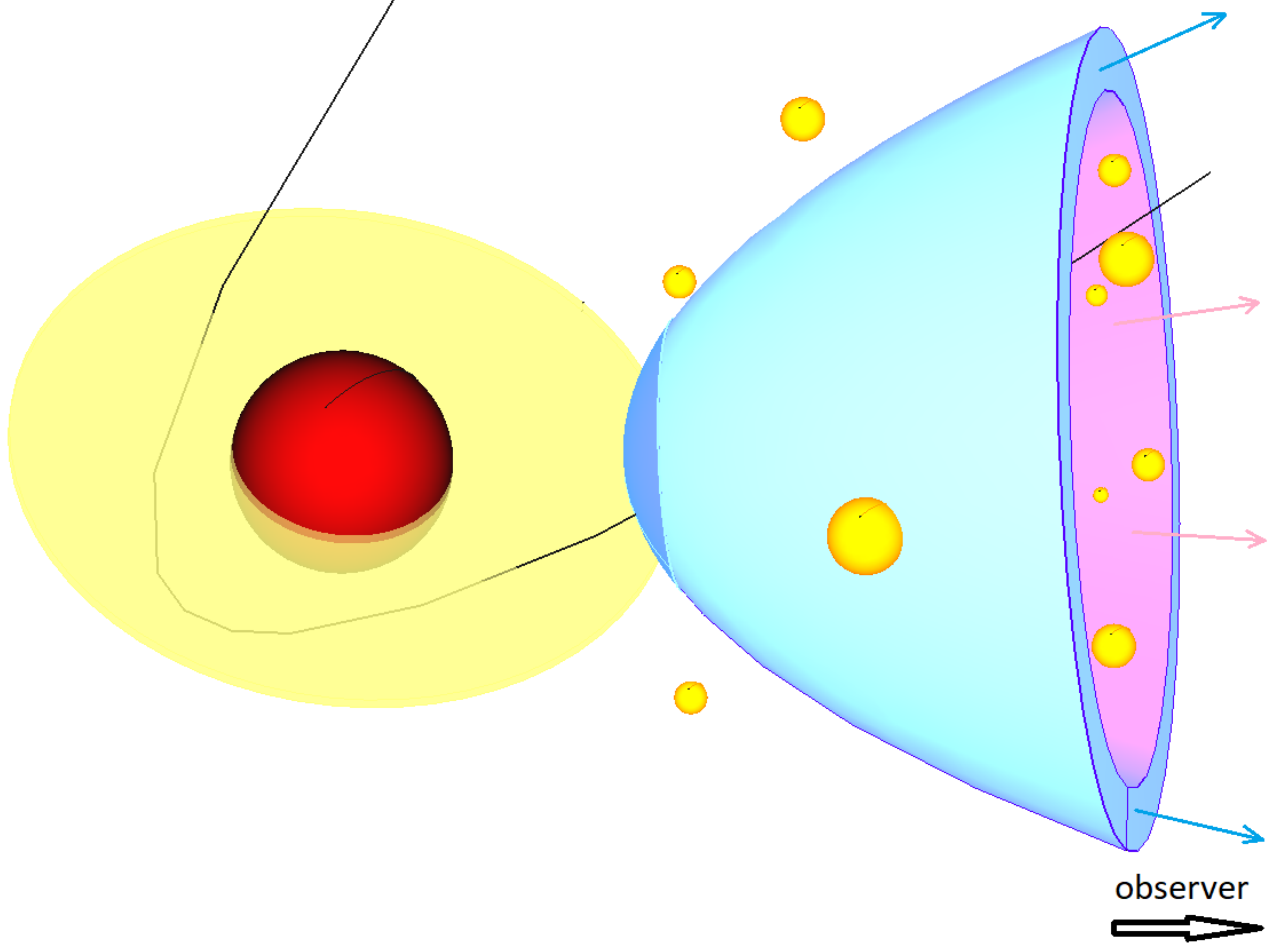}
\caption{A sketch of the geometry of the proposed model (not to the scale) for the period of GeV flare. Red sphere presents the Be star with the disk shown with yellow semi-transparent circle. Stellar/pulsar winds interaction cone is shown with cyan. \psrb pulsar orbit is illustrated with black line. The unshocked pulsar electrons (magenta) are strongly accelerated in the region close to the cone's tip (blue region; ``strongly shocked electrons'') and weakly accelerated at the rest of the cone surface(cyan region; ``weakly shocked electrons''). The flight directions of these electrons are shown with cyan/magenta arrows. The clumps of stellar wind are shown with yellow spheres. See text for the detailed model description.}
\label{fig:model_sketch}
\end{figure}

\begin{figure}
\includegraphics[width=\linewidth]{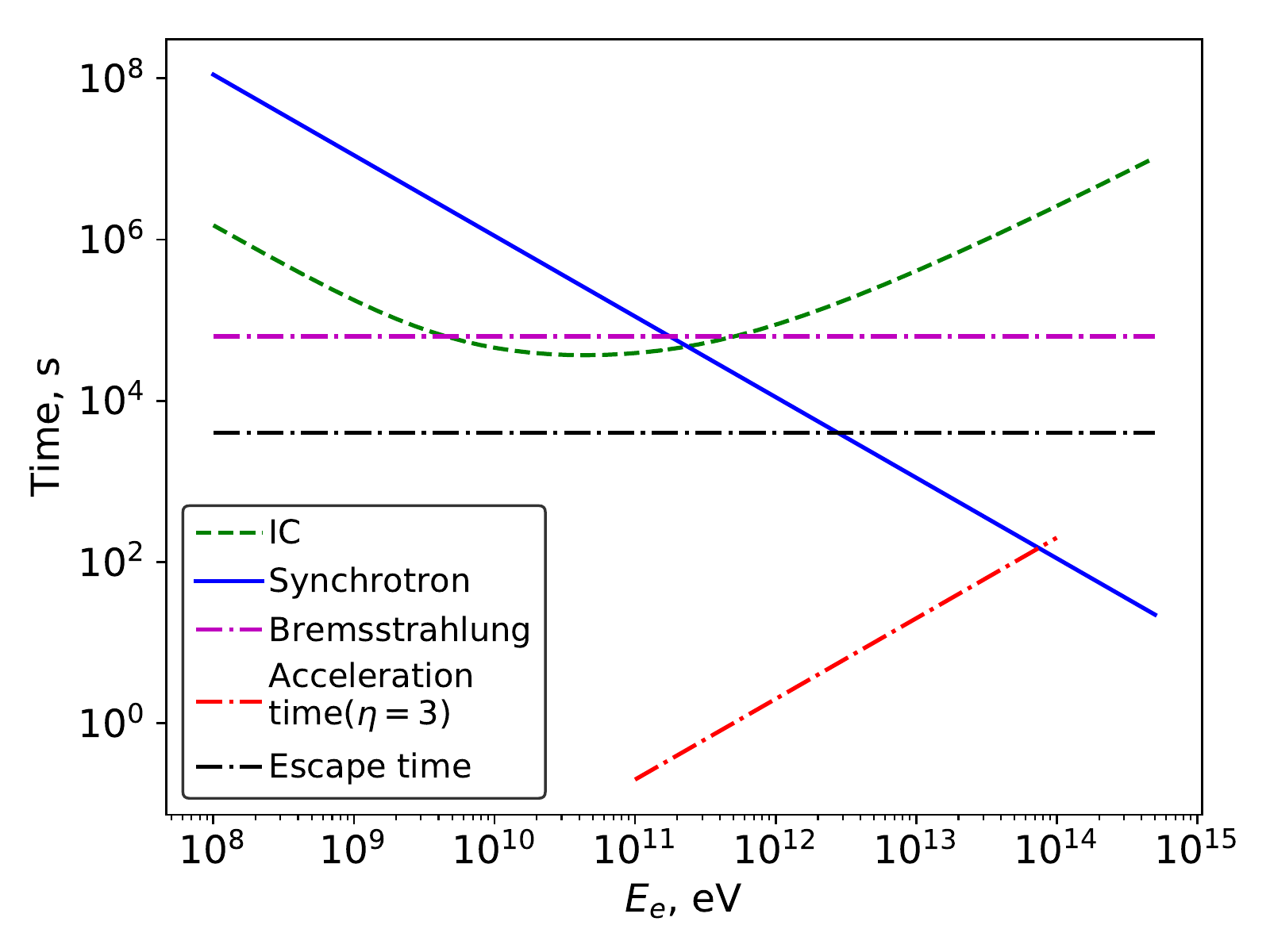}
\caption{Comparison of the cooling times to acceleration and escape time for various radiation processes in the case of avfl model, see text for details.}
\label{fig:losses}
\end{figure}

 The rest of the electrons flying into the shock direction will be reverted to flow along the shock at the surface of stellar-pulsar wind interaction cone far from the apex, and could be additionally mildly accelerated on a weak shock, see Fig.~\ref{fig:model_sketch} for a sketch of the model. This leads to a power law tail in the spectrum of diverted electrons with a slope $\sim -3$ which continues above $1$~GeV to at least $E_e\sim 5$~GeV. This slope is characteristic of particles acceleration on weak shocks~\citep[see e.g.][]{bell78,blandford87}. Hereafter we will refer to these diverted electrons as weakly shocked electrons.

 The spectra of both populations will be additionally modified by radiative (IC, synchrotron or bremsstrahlung) and non-radiative (adiabatic or escape) losses operating in the system. In our calculations the resulting electron spectrum was determined by numerically calculating the radiative losses of a continuously injected spectrum of electrons, until a steady solution has been obtained.
 The time that electrons spend in the emitting region, $R/c$, is about a few thousand seconds, and is long enough to substantially modify the injected spectrum due to synchrotron losses (in our calculations we took  t$_{esc}$=4000~s; see Figure~\ref{fig:losses} and Table~\ref{tab_mods}).

 We would like to note that the losses substantially modify the injected spectrum and thus properly accounting for  such losses is important for modelling the \psrb system.
 Additionally one has to account for the radiation efficiency of the considered mechanisms. This is rather low for the majority of the considered processes and subsequently an increase of the total energy of the pulsar wind is required (in comparison to 100\% efficiency case).
 
 According to our model, similar to previous works  \citep[e.g.][]{2014MNRAS.439..432C, chernyakova15, Chen19}, the  X-ray and TeV components are explained as synchrotron and IC emission of the strongly shocked electrons. The GeV component in our model is a separate component due to the combination of the IC and bremsstrahlung emission of the less energetic, unshocked electrons interacting  with the clumps of the matter from the Be-star wind/disk which penetrated through the shock.
 
\begin{figure*}
\includegraphics[width=\columnwidth]{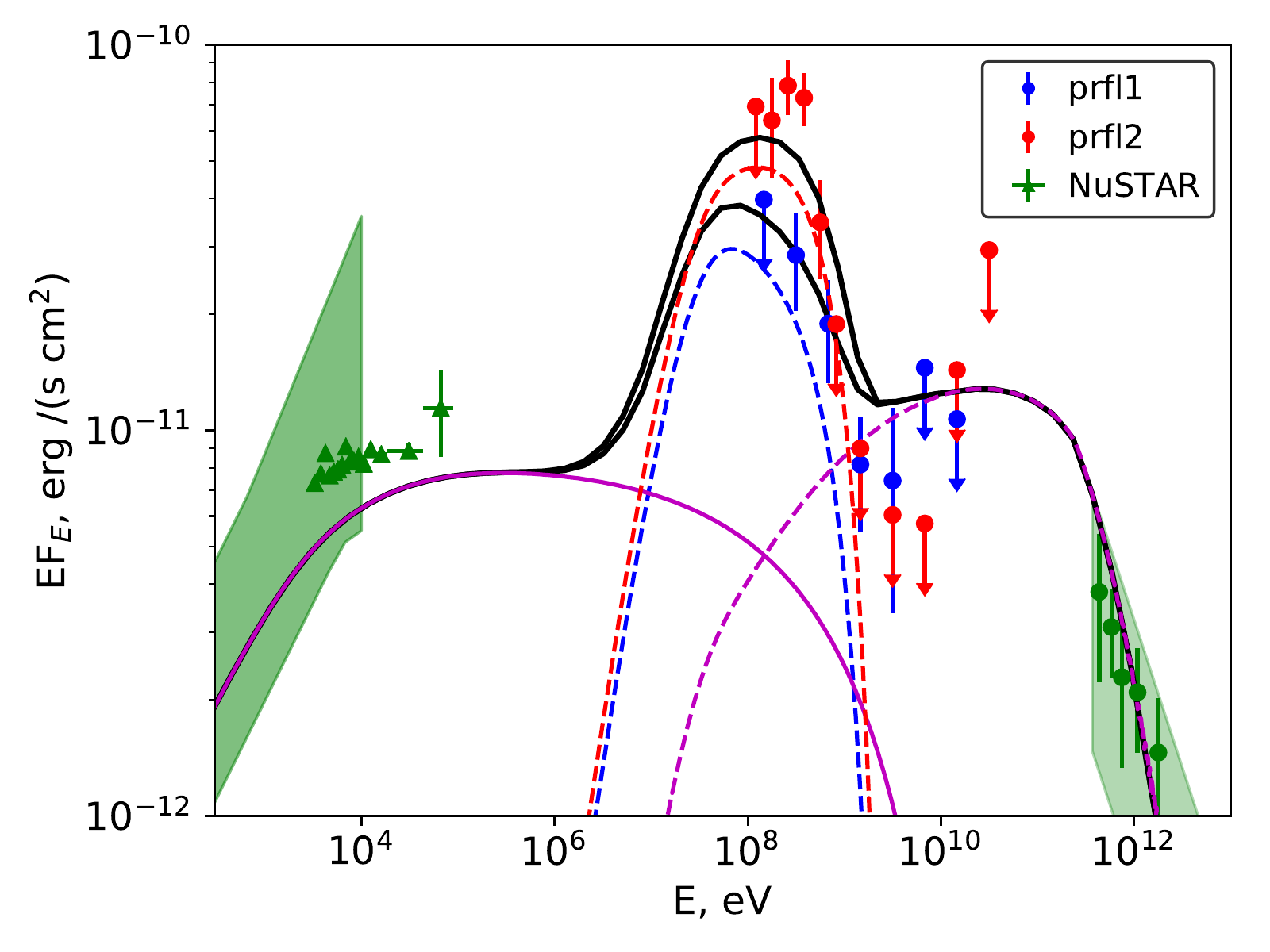}
\includegraphics[width=\columnwidth]{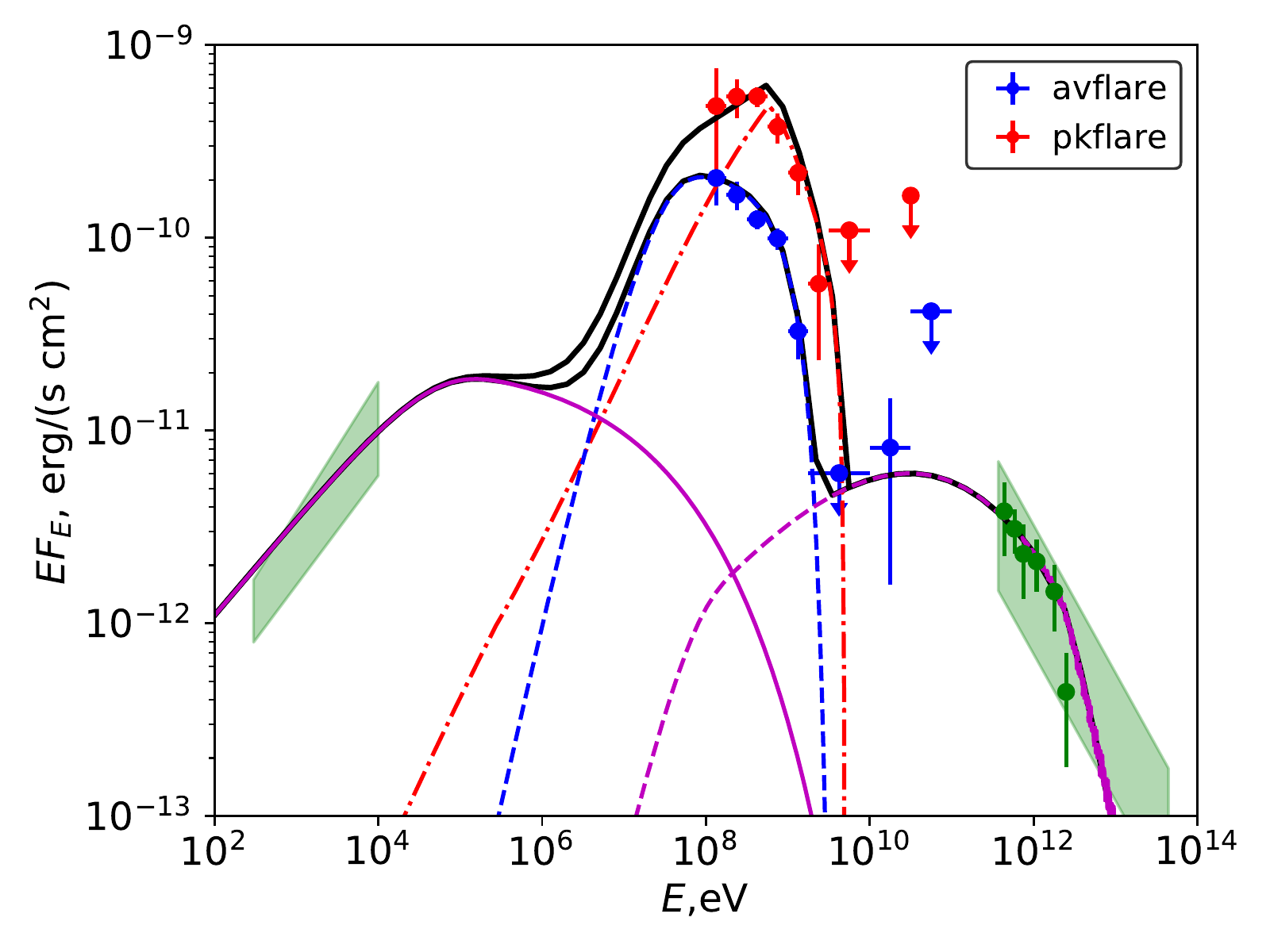}\caption{\textit{Left:} Broad band spectrum emission of PSR B1259-63 during the  20 days before (blue points) and after (red points) the periastron period. Green X-ray points are NuSTAR observations of the 2014 periastron from \citep{chernyakova15} 
\textit{Right:} Broad band spectrum emission of PSR B1259-63 during the  GeV flare. Blue points represent the flare averaged over the whole duration, and red points correspond to the sum of peak periods.
\newline  The TeV points shown are taken from  \citet{2005A&A...442....1A}, 
and the shaded regions at TeV energies represent the range of multi-years H.E.S.S. measurements reported in Fig. 2 of \citet{hess_psrb2020}.  
The shaded regions at X-ray energies represent the range of fluxes observed by SWIFT in 2017 before (left panel), and after (right panel) the GeV flare. In both panels dashed lines show an IC component, solid magenta line corresponds to synchrotron emission of strongly shocked electrons, dash-dotted line shows the bremsstrahlung component and black solid line corresponds to overall model emission.
}
\label{fig:avflare}
\end{figure*}

 The modelled SEDs along with the observed keV-TeV data are shown in Figure \ref{fig:avflare}.  In this figure the  synchrotron and IC emission of the strongly shocked electrons are shown with solid and dashed magenta curves correspondingly. The contributions of IC and bremsstrahlung emission of the unshocked and weakly shocked electrons are indicated  with dashed and dashed-dot curves correspondingly, where the colours refer to the prfl1 (blue) and prfl2 (red) periods. The TeV points shown are taken from \citet{2005A&A...442....1A}, and the coloured region at TeV energies represent the range of multi-years H.E.S.S. measurements reported in Fig. 2 of \citet{hess_psrb2020}.  The shaded region at X-ray energies represents the range of fluxes observed by \swift in 2017 before (left panel), and after (right panel) the GeV flare.  
 
 The synchrotron, bremsstrahlung and IC emission was calculated with the \texttt{naima v.0.8.3} package \citep{naima}, which uses the approximations for the IC, synchrotron and bremsstrahlung emission from ~\citet{aharonian81,aharonian10,khangulian14, bremss99}. 
 
 \begin{table}
\caption{Details of the models. D is a distance from the Be star to the emission region. Effective luminosity L of the pulsar wind electrons (without considering  beaming effects) is measured in units of spin-down luminosity L$_{sd}=8.2\times10^{35}$ erg/s}.
\label{tab_mods}
\begin{center}
\begin{small}
	\begin{tabular}{|c|c|c|c|c|c|c|}
 \hline
	Data & t-t$_p$,&D,   &   n$_{clump}$,   &B,  &$\Gamma_2$ & L/ L$_{sd}$         \\ 
	Period &days&10$^{13}$cm& 10$^{10}$cm$^{-3}$& G& & \\\hline
	fl15& &7.5&40&0.1&3&30\\\hline
	pkfl&41,56,58,70 &7.5&2&0.1&3&30\\\hline
	avfl&40 -- 72 &7.5&$\lesssim$ 0.1&0.1&3&30\\\hline
	prfl2&0 -- 20 & 2.5&$\lesssim$ 0.1&0.4& 2.5& 1 \\\hline
	prfl1&-20 -- 0 & 2.5&$\lesssim$ 0.1&0.4  & 3&1 \\\hline

\end{tabular}
\end{small}
\end{center}

\end{table}

 To explain the observed luminosity one  has to consider that for relativistic electrons both IC and bremsstrahlung radiation is strongly peaked in the forward direction and most of the energy radiated as a photon moving in the same direction as the initial electron. To explain the observed excess of the energy released during the short flares an initially isotropic pulsar wind should be reversed after the shock and confined within a cone, pointing in the direction of the observer, similar to the geometry described in~\citet{Khan11}.
 
 The difference in true anomaly between days 40 and 80 after the periastron (period including all short flares) is about 20$^\circ$. This angle is comfortably smaller than the apex angle of the shock $4\pi/30$ needed to explain the observed luminosity during the short flares in the case of 100\% efficiency. The cone of such a size is a result of the interaction of the isotropic pulsar wind with the isotropic Be star wind if the winds ram pressure ratio is $\frac{L_{sd}}{\dot{M}V_wc}=0.05$ \citep{Khan11}.

 In Figure \ref{fig:avflare} we show \flat spectra averaged over the total flare (green points), and over the peak of the flares only (red points). These spectra can be explained as a combination of IC and bremstrahlung emission on the clumps with  densities of  about $2\times10^{10}$ cm$^{-3}$ and less then $1\times10^{9}$ cm$^{-3}$ correspondingly, see Table \ref{tab_mods}. Please note that  in the Figure \ref{fig:avflare} we show only the dominant component (IC for the average flare and bremssrahlung for the peaks) in order to make the figure more readable. 
 Also please note that the split of the IC and bremsstrahlung is very model dependent and requires detailed hydrodynamic simulations beyond the scope of this paper. To explain the observed luminosity of the 15 minutes long flare one needs to assume a 1000 second long interaction with a clump  of material with a density of $4\times 10^{11}$ cm$^{-3}$.
 
 The required densities of the clumps are higher than  the average density of an undisturbed disc. At the same time the required averaged density both around the periastron and during the period of GeV flare is consistent with the unperturbed, smoothly decreasing disc density model in \citet{vanSoelen12}, which gives a density at the base of disc of $n_e \approx 6\times10^{13}$\,cm$^{-3}$. At the binary separation distance at 40 days from periastron the  disc density will have decreased to $\sim 10^8$\,cm$^{-3}$ (within the disc). This, combined with the observed H\,$\alpha$ variation (Fig.~\ref{fig:optical}), clearly indicates that the disc must be strongly clumped and disrupted near periastron. The difference in the 2017 \flat light curve (rapid flares) from the 2010 and 2014 periastra, also suggests a more complicated disc behaviour, that was unfortunately not observable during the 2017 periastron \citep[see also the discussion in][]{alma20}.

Recently a very hard TeV spectrum with the slope reaching values of $\Gamma\sim 2.5$ at certain orbital phases was reported in~\citet{hess_psrb2020}, see shaded region in Fig.~\ref{fig:avflare}. Assuming that this emission is produced by IC in a strong Klein-Nishina regime the spectral slope 
of the corresponding electrons can be estimated to be $\Gamma_e\sim 2.5$, similar to the value used in the modelling presented here.

In the case that the electrons propagate through regions with a non-zero magnetic field their spectrum, at TeV energies, undergoes severe cooling due to synchrotron losses , see e.g. Fig.~\ref{fig:losses}. 
This leads to the formation of a break in the electron spectrum with a typical softening after the break of $\Delta\Gamma_e=1$. Thus, to match the observed HESS spectrum, an initially extremely hard spectrum of electrons with a slope $\Gamma_{e,0}=1.5$ with a break to $\Gamma_e=2.5$ at TeV energies has to be considered.

The initial slope $\Gamma_{e,0}=1.5$ corresponds in the proposed model to the spectrum of the shocked electrons. This slope is substantially different from a ``standard'' $\Gamma_e=2$ slope of Fermi-mechanism accelerated electrons. We would like to note, however, that very hard slopes up to $\Gamma\sim 1$ (at least at relatively broad energy range close to the spectral cut-off) were reported for diffusive shock acceleration on a multiple shocks, see e.g.~\citet{melrose93,bykov13,vieu20}. Such shocks can potentially form in \psrb in a pulsar wind/Be star disc interaction region, assuming that the disc hosts multiple clumps.

GeV emission around the periastron period can be explained as IC emission of the 
unshocked electrons (see Figure \ref{fig:avflare}).  In this Figure we also show NuSTAR data points taken around 2014 periastron \citep{chernyakova15}. Within our model softening of the X-ray slope around the periastron can be attributed to additional cooling losses due to the higher value of magnetic field and the increased photon energy density 
near the periastron. 
To explain higher intensity and much sharper cutoff of the GeV flux after the periastron,  one needs 
to assume that the acceleration becomes more efficient and the slope of the electrons is equal to -2.5 above 1 GeV. 
The parameters of all the models are summarised in Table \ref{tab_mods}.

\section{Conclusions}
\label{sec:conclusions}
In this paper we present optical observations (spectroscopy and H\,$\alpha$ photometry) and discuss spectral characteristics of the GeV emission both around the periastron and during the flare. We propose a new model to explain the origin of the short bright GeV flares observed during 2017 periastron passage by Fermi. We show that:

1) The optical observations show that the behaviour of the disc around the first disc crossing is similar to the previous periastron, clearly indicating the pulsar significantly disrupts the disc. 

2) The observed X-ray and TeV emission both around the periastron and during GeV flare can be explained as a synchrotron and IC emission of the strongly shocked electrons of the pulsar wind. 

3) The GeV component is a combination of the IC emission of unshocked/ weakly shocked electrons and bremsstrahlung emission. 

4) The luminosity of the  GeV flares can be  understood if it is assumed that the initially isotropic pulsar wind after the shock is reversed and confined within a cone looking,  during the flare,  in the direction of the observer.

5) The observed softening of the spectrum close to the periastron corresponds to the shift of the break in the electrons spectrum due to the cooling losses. The position of the break is determined by the strength of the magnetic field in the emitting region, which is higher around the periastron. We foresee that the break position can be located at higher energies further from the periastron, which can lead to the detectable break in X-ray/TeV spectra. A hint of such a break was detected by Suzaku in 2007 \citep{Uchiyama2007}.

\section*{Acknowledgements} Authors thank Prof. F. Aharonian for fruitful discussions.
This paper uses observations made at the South African Astronomical Observatory (SAAO). This paper uses observations obtained at the Boyden Observatory, University of the Free State, South Africa. Watcher data made available through support from Science Foundation Ireland grant 07/RFP/PHYF295.  The authors acknowledge support by the state of Baden-W\"urttemberg through bwHPC. This work was supported by DFG through the grant MA 7807/2-1. The authors wish to acknowledge the DJEI/DES/SFI/HEA Irish Centre for High-End Computing (ICHEC) for the provision of computational facilities and support. We would also like to acknowledge networking support by the COST Actions CA16214 and CA16104. DM acknowledges support from the Irish Research Council through grant GOIPG/2014/453.

\section*{Data availability}
Optical data underlying this article were provided  by permission of SALT/Watcher collaborations. Data will be shared on request to the corresponding author with the permission of SALT/Watcher collaborations. Other data used in the article will be shared on reasonable request to the corresponding author.

\def\aj{AJ}%
\def\actaa{Acta Astron.}%
\def\araa{ARA\&A}%
\def\apj{ApJ}%
\def\apjl{ApJ}%
\def\apjs{ApJS}%
\def\ao{Appl.~Opt.}%
\def\apss{Ap\&SS}%
\def\aap{A\&A}%
\def\aapr{A\&A~Rev.}%
\def\aaps{A\&AS}%
\def\azh{AZh}%
\def\baas{BAAS}%
\def\bac{Bull. astr. Inst. Czechosl.}%
\def\caa{Chinese Astron. Astrophys.}%
\def\cjaa{Chinese J. Astron. Astrophys.}%
\def\icarus{Icarus}%
\def\jcap{J. Cosmology Astropart. Phys.}%
\def\jrasc{JRASC}%
\def\mnras{MNRAS}%
\def\memras{MmRAS}%
\def\na{New A}%
\def\nar{New A Rev.}%
\def\pasa{PASA}%
\def\pra{Phys.~Rev.~A}%
\def\prb{Phys.~Rev.~B}%
\def\prc{Phys.~Rev.~C}%
\def\prd{Phys.~Rev.~D}%
\def\pre{Phys.~Rev.~E}%
\def\prl{Phys.~Rev.~Lett.}%
\def\pasp{PASP}%
\def\pasj{PASJ}%
\def\qjras{QJRAS}%
\def\rmxaa{Rev. Mexicana Astron. Astrofis.}%
\def\skytel{S\&T}%
\def\solphys{Sol.~Phys.}%
\def\sovast{Soviet~Ast.}%
\def\ssr{Space~Sci.~Rev.}%
\def\zap{ZAp}%
\def\nat{Nature}%
\def\iaucirc{IAU~Circ.}%
\def\aplett{Astrophys.~Lett.}%
\def\apspr{Astrophys.~Space~Phys.~Res.}%
\def\bain{Bull.~Astron.~Inst.~Netherlands}%
\def\fcp{Fund.~Cosmic~Phys.}%
\def\gca{Geochim.~Cosmochim.~Acta}%
\def\grl{Geophys.~Res.~Lett.}%
\def\jcp{J.~Chem.~Phys.}%
\def\jgr{J.~Geophys.~Res.}%
\def\jqsrt{J.~Quant.~Spec.~Radiat.~Transf.}%
\def\memsai{Mem.~Soc.~Astron.~Italiana}%
\def\nphysa{Nucl.~Phys.~A}%
\def\physrep{Phys.~Rep.}%
\def\physscr{Phys.~Scr}%
\def\planss{Planet.~Space~Sci.}%
\def\procspie{Proc.~SPIE}%
\let\astap=\aap
\let\apjlett=\apjl
\let\apjsupp=\apjs
\let\applopt=\ao

\bibliography{references}
\label{lastpage}
\end{document}